\documentclass[onecolumn]{aa}

\usepackage{graphicx}
\usepackage{natbib}
\usepackage{amsmath}
\bibpunct{(}{)}{;}{a}{}{,}

\newcommand{\twofig}[4]{%
  \begin{figure*}%
    \centerline{\resizebox{\hsize}{!}{\includegraphics*{#1} \,%
                                      \includegraphics*{#2}}}%
    \caption{#4}\label{#3}%
  \end{figure*}%
}
\newcommand{\threefig}[5]{%
  \begin{figure*}%
    \centerline{\resizebox{\hsize}{!}{\includegraphics*{#1} \,%
                                      \includegraphics*{#2} \,%
                                      \includegraphics*{#3}}}%
    \caption{#5}\label{#4}%
  \end{figure*}%
}

\newcommand{\sect}[1]{Sect.~\ref{#1}}
\newcommand{\fig}[1]{Fig.~\ref{#1}}
\newcommand{\eq}[1]{Eq.~\ref{#1}}
\newcommand{\eqs}[1]{Eqs.~\ref{#1}}

\newcommand{\gn}{Gaussian}
\newcommand{\ngn}{non-Gaussian}
\newcommand{\ngy}{non-Gaussianity}
\newcommand{\ngs}{non-Gaussianities}
\newcommand{\LL}{{\mathrm{L}}}
\newcommand{\NL}{{\mathrm{NL}}}
\newcommand{\fnl}{$f_{\NL}$}

\begin{document}

\title{Probing local non-Gaussianities within a Bayesian framework}

\titlerunning{Probing local non-Gaussianities}

\author{Franz Elsner\inst{1},
  Benjamin D. Wandelt\inst{2,3}, \and
  Michael D. Schneider\inst{4}}

\authorrunning{Elsner et al.}

\offprints{felsner@mpa-garching.mpg.de}

\institute{Max-Planck-Institut f\"ur Astrophysik,
Karl-Schwarzschild-Stra\ss e 1, D-85748 Garching, Germany
\and
Department of Physics, University of Illinois at
Urbana-Champaign, 1110 W.~Green Street, Urbana, IL 61801, USA
\and
Department of Astronomy, University of Illinois at
Urbana-Champaign, 1002 W.~Green Street, Urbana, IL 61801, USA
\and
Institute for Computational Cosmology, Department of Physics, Durham
  University, South Road, Durham, DH1 3LE, UK}

\date{Received \dots / Accepted \dots}

\abstract
{}
{We outline the Bayesian approach to inferring \fnl, the level of
  \ngs\ of local type. Phrasing \fnl\ inference in a Bayesian
  framework takes advantage of existing techniques to account for
  instrumental effects and foreground contamination in CMB data and
  takes into account uncertainties in the cosmological parameters in
  an unambiguous way.}
{We derive closed form expressions for the joint posterior of \fnl\
  and the reconstructed underlying curvature perturbation, $\Phi$, and
  deduce the conditional probability densities for \fnl\ and
  $\Phi$. Completing the inference problem amounts to finding the
  marginal density for \fnl. For realistic data sets the necessary
  integrations are intractable. We propose an exact Hamiltonian
  sampling algorithm to generate correlated samples from the \fnl\
  posterior. For sufficiently high signal-to-noise ratios, we can
  exploit the assumption of weak \ngy\ to find a direct Monte Carlo
  technique to generate \emph{independent} samples from the posterior
  distribution for \fnl. We illustrate our approach using a simplified
  toy model of CMB data for the simple case of a 1-D sky.}
{When applied to our toy problem, we find that, in the limit of high
  signal-to-noise, the sampling efficiency of the approximate
  algorithm outperforms that of Hamiltonian sampling by two orders of
  magnitude. When \fnl\ is not significantly constrained by the data,
  the more efficient, approximate algorithm biases the posterior
  density towards $f_{\NL} = 0$.}
{}


\maketitle
%________________________________________________________________
%________________________________________________________________
%________________________________________________________________

\section{Introduction}
\label{sec:intro}

The analysis of cosmic microwave background (CMB) radiation data has
considerably improved our understanding of cosmology and played a
crucial role in constraining the set of fundamental cosmological
parameters of the universe \citep{2007ApJS..170..377S,
  2009ApJS..180..225H}. This success is based on the intimate link
between the temperature fluctuations we observe today and the physical
processes taking place in the very early universe. Inflation is
currently the favored theory predicting the shape of primordial
perturbations \citep{1981PhRvD..23..347G, 1982PhLB..108..389L}, which
in its canonical form leads to very small \ngs\ that are far from
being detectable by means of present-day experiments
\citep{Maldacena:2002vr, Acquaviva:2002ud}. However, inflation
scenarios producing larger amounts of \ngy\ can naturally be
constructed by breaking one or more of the following properties of
canonical inflation: slow-roll, single-field, Bunch-Davies vacuum, or
a canonical kinetic term \citep{2004PhR...402..103B}. Thus, a positive
detection of primordial \ngy\ would allow us to rule out the simplest
models. Combined with improving constraints on the scalar spectral
index $n_s$, the test for \ngy\ is therefore complementary to the
search for gravitational waves as a means to test the physics of the
early Universe.

A common strategy for estimating primordial \ngy\ is to examine a
cubic combinations of filtered CMB sky maps
\citep{2005ApJ...634...14K}. This approach takes advantage of the
specific bispectrum signatures produced by primordial \ngy\ and yields
to a computationally efficient algorithm. When combined with the
variance reduction technique first described by
\citet{2006JCAP...05..004C} these bispectrum-based techniques are close
to optimal, where optimality is defined as saturation of the
Cramer-Rao bound. Lately, a more computationally costly minimum
variance estimator has been implemented and applied to the WMAP5 data
\citep{2009JCAP...09..006S}.

Recently, a Bayesian approach has been introduced in CMB power
spectrum analysis and applied successfully to WMAP data making use of
Gibbs-sampling techniques \citep{2004ApJ...609....1J,
  2004PhRvD..70h3511W}. Within this framework, one draws samples from
the posterior probability density given the data without explicitly
calculating it. The target probability distribution is finally
constructed out of the samples directly, thus computationally costly
evaluations of the likelihood function or its derivatives are not
necessary. Another advantage of the Bayesian analysis is that the
method naturally offers the possibility to include a consistent
treatment of the uncertainties associated with foreground emission or
instrumental effects \citep{2008ApJ...672L..87E}. As it is possible to
model CMB and foregrounds jointly, statistical interdependencies can
be directly factored into the calculations. This is not
straightforward in the frequentist approach where the data analysis is
usually performed in consecutive steps. Yet another important and
desirable feature is the fact that a Bayesian analysis obviates the
necessity to specify fiducial parameters, whereas in the frequentist
approach it is only possible to test one individual null hypothesis at
a time.

In this paper, we pursue the modest goal of developing the formalism
for the extension of the Bayesian approach to the analysis of \ngn\
signals, in particular to local models, where the primordial
perturbations can be modeled as a spatially local, non-linear
transformation of a \gn\ random field. Utilizing this method, we are
able to write down the full posterior probability density function
(PDF) of the level of \ngy. We demonstrate the principal aspects of
our approach using a 1-D toy sky model. Although we draw our
discussion on the example of CMB data analysis, the formalism
presented here is of general validity and may also be applied within a
different context.

The paper is organized as follows. In \sect{sec:basics} we give a
short overview of the theoretical background used to characterize
primordial perturbations. We present a new approximative approach to
extract the amplitude of \ngs\ from a map in \sect{sec:sampling} and
verify the method by means of a simple synthetic data model
(\sect{sec:model}). We compare the performance of our technique to an
exact Hamiltonian Monte Carlo sampler which we developer in
\sect{sec:hmc} and discuss the extensions of the model required to
deal with a realistic CMB sky map (\sect{sec:cmb}). Finally, we
summarize our results in \sect{sec:summary}.

\section{Model of non-Gaussianity}
\label{sec:basics}

The expansion coefficients $a_{\ell m}$ of the observed CMB
temperature anisotropies in harmonic space can be related to the
primordial fluctuations via
\begin{equation}
\label{eq:alm2phi}
a_{\ell m}=\frac{2b_{\ell}}{\pi}\int k^2 dk \, r^2 dr \, \lbrack \,
\Phi_{\ell m}(r) \, g^{adi}_{\ell}(k) + S_{\ell m}(r) \,
g^{iso}_{\ell}(k) \, \rbrack \, j_\ell(kr) + n_{\ell m} \, ,
\end{equation}
where $\Phi_{\ell m}(r)$ and $S_{\ell m}(r)$ are the primordial
curvature and isocurvature perturbations at comoving distance $r$,
$g^{adi}_{\ell}(k)$ and $g^{iso}_{\ell}(k)$ their corresponding
transfer functions in momentum space. The spherical Bessel function of
order $\ell$ is denoted by $j_\ell(kr)$, $b_{\ell}$ includes beam
smearing effects, and $n_{\ell m}$ describes instrumental noise. As
curvature perturbations dominate over isocurvature perturbations
\citep{2006PhRvD..74f3503B, 2007MNRAS.375L..26T}, we will neglect the
contribution of $S_{\ell m}$ in our subsequent analysis.

Any \ngn\ signature imprinted in the primordial perturbations will be
transferred to the $a_{\ell m}$ according to \eq{eq:alm2phi} and is
therefore detectable, in principle. Theoretical models predicting
significant levels of \ngn\ contributions to the observed signal can
be subdivided into two broad classes \citep{2004JCAP...08..009B}: one
producing \ngy\ of \emph{local} type, the other of \emph{equilateral}
type. The former kind of \ngy\ is achieved to very good approximation
in multi-field inflation as described by the curvaton model
\citep{Moroi:2001ct, Enqvist:2001zp, 2003PhRvD..67b3503L}, or in
cyclic/ekpyrotic universe models \citep{2001PhRvD..64l3522K,
  2002PhRvD..65l6003S}. The latter type of \ngy\ is typically a result
of single field models with non-minimal Lagrangian including higher
order derivatives \citep{2004PhRvD..70l3505A, 2005PhRvD..71d3512S}.

Concentrating on \emph{local} models, we can parametrize the \ngy\ of
$\Phi$ by introducing an additional quadratic dependence on a purely
\gn\ auxiliary field $\Phi_{\LL}$, that is local in real space, of the
form \citep{2000MNRAS.313..141V, 2001PhRvD..63f3002K}
\begin{equation}
\label{eq:fnldef}
\Phi_{\NL}(r)=\Phi_{\LL}(r) + f_{\NL}\lbrack \Phi^2_{\LL}(r)-\langle
\Phi^2_{\LL}(r) \rangle \rbrack \, ,
\end{equation}
where \fnl\ is a dimensionless measure of the amplitude of \ngy.

\section{Bayesian inference of non-Gaussianity}
\label{sec:sampling}

It has been shown to be feasible to reconstruct the primordial
curvature potential out of temperature or temperature and polarization
sky maps \citep{2005PhRvD..71l3004Y, 2009ApJS..184..264E}, which allows
searching for primordial \ngs\ more sensitively. Although the mapping
from a 3D potential to a 2D CMB sky map is not invertible
unambiguously, a unique solution can be found by requiring that the
result minimizes the variance. In this conventional frequentist
approach, the level of \ngy\ and an estimate of its error is derived
from a cubic combination of filtered sky maps
\citep{2005ApJ...634...14K}. We will show in the following sections
how to sample \fnl\ from the data and unveil the full posterior PDF
using a Bayesian approach.

\subsection{Joint probability distribution}

In our analysis we assume the data vector $d$ to be a superposition of
the CMB signal $s$ and additive noise $n$
\begin{align}
d &= B s + n \nonumber \\ &= B M \Phi + n \, ,
\end{align}
where information about observing strategy and the optical system are
encoded in a pointing matrix $B$ and $M$ is a linear transformation
matrix. In harmonic space, the signal is related to the primordial
scalar perturbation as
\begin{align}
s_{\ell m} &= \frac{2}{\pi} \int k^2 dk \, r^2 dr\, \Phi_{\ell m}(r)
\, g^{adi}_{\ell}(k) \, j_\ell(kr) \nonumber \\ &\approx \sum_{i}
M_{i} \Phi_{\ell m}(r_i) \nonumber \\ &\equiv M \Phi_{\ell m} \, .
\end{align}

Our aim is to construct the posterior PDF of the amplitude of \ngs\
given the data, $P(f_{\NL} | d)$. To do so, we subsume the remaining
set of cosmological parameters to a vector $\theta$ and calculate the
joint distribution as
\begin{equation}
P(d,\Phi_{\LL},f_{\NL},\theta) = P(d|\Phi_{\LL},f_{\NL},\theta)
P(\Phi_{\LL}|\theta) P(\theta) P(f_{\NL}) \, .
\end{equation}
Now, we can use \eq{eq:fnldef} to express the probability for data $d$
given $\Phi_{\LL}$, \fnl, and $\theta$
\begin{equation}
  P(d|\Phi_{\LL},f_{\NL},\theta) = \frac{1}{\sqrt{|2 \pi N|}} \,
  \mathrm{e} \, ^{ {-1/2 \, \lbrack d - B M ( \Phi_{\LL} + f_{\NL}
      (\Phi^2_{\LL} - \langle \Phi^2_{\LL} \rangle)) \rbrack^{\dagger}
      N^{-1} \lbrack d - B M ( \Phi_{\LL} + f_{\NL} (\Phi^2_{\LL} -
      \langle \Phi^2_{\LL} \rangle) ) \rbrack } } \, ,
\end{equation}
where $N$ is the noise covariance matrix. The prior probability
distribution for $\Phi_{\LL}$ given $\theta$ can be expressed by a
multivariate Gaussian by definition. Using the covariance matrix
$P_\Phi$ of the potential, we derive
\begin{equation}
\label{eq:philprior}
P(\Phi_{\LL}|\theta) = \frac{1}{\sqrt{|2 \pi P_\Phi|}} \, \mathrm{e}
\, ^{ {-1/2 \, \Phi_{\LL}^{\dagger} P_\Phi^{-1} \Phi_{\LL}} } \, .
\end{equation}
For flat priors $P(f_{\NL})$, $P(\theta)$ we finally obtain
\begin{multline}
\label{eq:jointdistr_l}
P(d,\Phi_{\LL},f_{\NL},\theta) \propto \exp \left\{
  -\frac{1}{2}\, \left[ \vphantom{Phi_{\LL}^{\dagger}} (d - B M
    (\Phi_{\LL} + f_{\NL} (\Phi^2_{\LL} - \langle \Phi^2_{\LL}
    \rangle)))^{\dagger} N^{-1} \right. \right. \\
\left. \left. \times (d - B M ( \Phi_{\LL} + f_{\NL} (\Phi^2_{\LL} -
    \langle \Phi^2_{\LL} \rangle))) + \Phi_{\LL}^{\dagger}
    P^{-1}_{\Phi} \Phi_{\LL} \right] \vphantom{\frac{1}{2}} \right\}
\end{multline}
as an exact expression for the joint distribution up to a
normalization factor.

To derive the posterior density, $P(f_{\NL} | d)$, one has to
marginalize the joint distribution over $\Phi_{\LL}$ and $\theta$. As
it is not possible to calculate the high dimensional $\Phi_{\LL}$
integral directly, an effective sampling scheme must be found to
evaluate the expression by means of a Monte Carlo algorithm. One
possibility would be to let a Gibbs sampler explore the parameter
space. Unfortunately, we were not able to find an efficient sampling
recipe from the conditional densities for \fnl\ and $\Phi_{\LL}$ as
the variables are highly correlated. An algorithm that also generates
correlated samples, but is potentially suitable for \ngn\ densities
and high degrees of correlation is the Hamiltonian Monte Carlo
approach. We will return to this approach in \sect{sec:hmc}.

For now we attempt to go beyond correlated samplers and see whether we
can develop an approximate scheme, valid in the limit of weak \ngy, to
sample \fnl\ independently. We start out by expanding the target
posterior distribution into an integral of conditional probabilities
over the non-linear potential $\Phi_{\NL}$,
\begin{equation}
\label{eq:posterior}
P(f_{\NL}| d) = \int d\Phi_{\NL} d\theta \,
P(f_{\NL}|\Phi_{\NL},\theta) P(\Phi_{\NL}|d,\theta) P(\theta| d) \, .
\end{equation}
To construct the conditional probability $P(\Phi_{\NL}|d,\theta)$ in
the integrand, we need to find an equivalent equation for the joint
distribution (\eq{eq:jointdistr_l}) as a function of the field
$\Phi_{\NL}$. However, a simple analytic expression for the prior
distribution of $\Phi_{\NL}$ does not exist because it is a non-linear
transform of the \gn\ auxiliary field $\Phi_{\LL}$. To quantify the
expected correction, we calculate its covariance matrix,
\begin{align}
  (P_{\Phi_{\NL}})_{ij} &= \langle (\Phi_{\NL})_i (\Phi_{\NL})_j
  \rangle \nonumber \\ &= \langle (\Phi_{\LL})_i (\Phi_{\LL})_j
  \rangle + f_{\NL}^2 [ \langle (\Phi_{\LL})^2_i (\Phi_{\LL})^2_j
  \rangle - \langle (\Phi_{\LL})^2_i \rangle \langle (\Phi_{\LL})^2_j
  \rangle ] \nonumber \\ &= (P_{\Phi_{\LL}})_{ij} + 2 f_{\NL}^2
  (P_{\Phi_{\LL}})^2_{ij} \, .
\end{align}
As the covariance matrix $P_{\Phi_{\LL}}$ is of the order
$\mathcal{O}(10^{-10})$ and the \ngn\ contribution to $\Phi_{\NL}$ is
known to be small, we neglect the higher order correction in the prior
distribution in what follows. That is, we approximate the true prior
probability function by a \gn\ distribution in $\Phi_{\NL}$ and in
this way derive a simple expression for the joint density, as a
function of $\Phi_{\NL}$,
\begin{equation}
\label{eq:jointdistr_nl}
P(d,\Phi_{\NL},\theta) \propto \exp \left\{ -\frac{1}{2}\,
  \left[ (d - B M \Phi_{\NL})^{\dagger} N^{-1} (d - B M \Phi_{\NL}) +
    \Phi_{\NL}^{\dagger} P_{\Phi}^{-1} \Phi_{\NL} \right] \right\} \, .
\end{equation}
Note, that the approximation applies to the second term only, the
first part of the expression remains unaffected. As this approximation
is equivalent to imposing the prior belief of purely \gn\ primordial
perturbations, we expect to underestimate \fnl\ in the low
signal-to-noise regime, as we tend to replace the Wiener filtered
noise with purely \gn\ signal. Contrary, the method is unbiased when
the likelihood dominates over the prior which is unlikely for data
derived by the Planck mission.

The direct evaluation of the joint distributions over a grid in the
high dimensional parameter space is computationally not feasible. One
option would be to approximate the PDF around its maximum to get an
expression for the attributed errors \citep{1997PhRvD..55.5895T,
  1998PhRvD..57.2117B}. These methods are still computationally
expensive and can also not recover the full posterior. An alternative
approach to overcome these problems is to draw samples from the PDF
which is to be evaluated as we will discuss in the next section.

\subsection{Conditional probabilities}

To construct the target posterior density \eq{eq:posterior}, we have
to find expressions for the conditional probabilities
$P(\Phi_{\NL}|d,\theta)$ and $P(f_{\NL}|\Phi_{\NL},\theta)$. The
former distribution can easily be derived from the joint probability
density \eq{eq:jointdistr_nl}. Since the exponent is quadratic in
$\Phi_{\NL}$ in our approximation, the conditional PDF of $\Phi_{\NL}$
given $d$ and $\theta$ is \gn. Therefore, we can calculate mean and
variance of the distribution via differentiating the expression,
\begin{align}
\label{eq:phi_mean_var}
\langle \Phi_{\NL} \rangle &=  \langle (\Phi_{\NL} - \langle
\Phi_{\NL} \rangle )^2 \rangle \ M^{\dagger} B^{\dagger} N^{-1} d
\nonumber \\
\langle (\Phi_{\NL} - \langle \Phi_{\NL} \rangle )^2 \rangle &= \left[
  M^{\dagger} B^{\dagger} N^{-1} B M + P_\Phi^{-1} \right]^{-1} \, .
\end{align}
As a next step, we derive the conditional probability distribution of
\fnl\ for given $\Phi_{\NL}$. This expression is not affected by the
approximation and can be derived from a marginalization over
$\Phi_{\LL}$,
\begin{align}
P(f_{\NL} | \Phi_{\NL}, \theta) &= \int d\Phi_{\LL} \, P(f_{\NL} |
\Phi_{\LL}, \Phi_{\NL}) \, P(\Phi_{\LL} | \theta) \nonumber \\ &= \int
d\Phi_{\LL} \, \delta (\Phi_{\NL} - \Phi_{\LL} - f_{\NL}(\Phi_{\LL}^2
- \langle \Phi_{\LL}^2 \rangle)) \, P(\Phi_{\LL} | \theta) \, .
\end{align}
Using \eq{eq:philprior}, we can calculate the integral and obtain
\begin{equation}
\label{eq:condfnl}
P(f_{\NL} | \Phi_{\NL}, \theta) \propto \left| \prod_i \frac{1}{1 + 2
    f_{\NL} (\tilde{\Phi}_{\LL})_i} \right| \, \mathrm{e} \, ^{ {-1/2
    \, \tilde{\Phi}_{\LL}^{\dagger} P_\Phi^{-1} \tilde{\Phi}_{\LL}} } \, ,
\end{equation}
where $\tilde{\Phi}_{\LL}$ is a function of \fnl\ and can be regarded
as inversion of \eq{eq:fnldef},
\begin{equation}
\label{eq:invphinl}
\tilde{\Phi}_{\LL} = \frac{1}{2 f_{\NL}} \left[ -1 + \sqrt{1 + 4
    f_{\NL}(\Phi_{\NL} + f_{\NL} \langle \Phi^2_{\LL} \rangle)} \right] \, .
\end{equation}
Note that we can resolve the ambiguity in sign in the weakly \ngn\
limit \citep{2005PhRvD..72d3003B}. Because the absolute value of the
elements of the second solution is typically larger by orders of
magnitude, the probability of its realization is strongly disfavored
by the prior $P(\Phi_{\LL})$. The factor of suppression is typically
less than $10^{-1000}$ and further vanishing with decreasing \fnl.

After setting up the conditional densities, we now can sample from the
distributions iteratively. First, we draw $\Phi_{\NL}$ from a \gn\
distribution using \eqs{eq:phi_mean_var}. Then, \fnl\ can be sampled
according to \eq{eq:condfnl} using the value of $\Phi_{\NL}$ derived
in the preceding step. Thus, the sampling scheme reads as
\begin{align}
\label{eq:scheme}
\Phi_{\NL}^i &\hookleftarrow P(\Phi_{\NL} | d, \theta) \nonumber \\
f_{\NL}^i &\hookleftarrow P(f_{\NL}| \Phi_{\NL}^i, \theta) \, .
\end{align}

Note that this is \emph{not} Gibbs sampling. For a fixed set of
cosmological parameters, we can chain together samples from the
conditional densities above, producing \emph{independent} \fnl\
samples. The efficiency of such a direct Monte Carlo sampler is
therefore expected to be much higher than that of a Gibbs sampler,
which, in the general case, would produce correlated samples.

As an extension of the sampling scheme presented so far, we sketch an
approach to account for uncertainties in cosmological parameters and
foreground contributions. Complementing the scheme (\eqs{eq:scheme})
by an additional step allows to take into account the error in the
parameters $\theta$,
\begin{align}
  \Phi_{\NL}^i &\hookleftarrow P(\Phi_{\NL} | d, \theta^{i-1}) \nonumber \\
  f_{\NL}^i &\hookleftarrow P(f_{\NL}| \Phi_{\NL}^i, \theta^{i-1}) \nonumber \\
  \theta^i &\hookleftarrow P(\theta | d, f_{\NL}^i) \, ,
\end{align}
where the last equation updates the cosmological parameters that can
be sampled from the data by means of standard Monte Carlo analysis
tools\footnote{E.g.\ as described in
  \citet{2002PhRvD..66j3511L}}. Now, the scheme formally reads as a
Gibbs sampler and can in principle take into account the correlation
among \fnl\ and the other cosmological parameters exactly. In
practice, however, the impact of a non-vanishing \fnl\ is expected to
be negligible, i.e.\ $P(\theta | d, f_{\NL}) \approx P(\theta |
d)$. Likewise, we can allow for an additional sampling step to deal
with foreground contributions, e.g.\ from synchrotron, free-free, and
dust emission. Foreground templates $f^{sync, \, free, \, dust}$, that
are available for these sources, can be subtracted with amplitudes
$c^{sync, \, free, \, dust}$ which are sampled from the data in each
iteration, $c^i \hookleftarrow P(c | d, f^{sync, \, free, \, dust},
\theta^i)$ \citep{2004PhRvD..70h3511W}. Alternatively, component
separation techniques could be used to take foreground contaminants
into account without the need to rely on a priori defined templates
\citep{2006ApJ...641..665E}. The traditional approach to deal with
point sources is to mask affected regions of the sky to exclude them
from the analysis. Discrete object detection has been demonstrated to
be possible within a Bayesian framework \citep{2003MNRAS.338..765H,
  2009MNRAS.393..681C}, and can be fully included into the sampling
chain. However, as sources are only successfully detected down to an
experiment-specific flux limit, a residue-free removal of their
contribution is in general not possible.

As the angular resolution of sky maps produced by existing CMB
experiments like WMAP is high and will further increase once data of
the Planck satellite mission becomes available, computational
feasibility of an analysis tool is an issue. The speed of our method
in a full implementation is limited by harmonic transforms which scale
as $\mathcal{O}(N_{pix}^{3/2})$ and are needed to calculate the
primordial perturbations at numerous shells at distances from the
cosmic horizon to zero. Thus, it shows the same scaling relation as
fast cubic estimators \citep{2005ApJ...634...14K,
  2007ApJ...664..680Y}, albeit with a larger prefactor.

\section{Implementation and Discussion}
\label{sec:model}

To verify our results and demonstrate the applicability of the method,
we implemented a simple 1-D toy model. We considered a vector
$\Phi_{\LL}$ of random numbers generated from a heptadiagonal
covariance matrix with elements
\begin{equation}
P_\Phi =
\begin{pmatrix}
& & & \ddots \\
\dots \, 0 & 0.1 & 0.2 & 0.5 & 1.0 & 0.5 & 0.2 & 0.1 & 0 \, \dots \\
& & & & & \ddots \\
\end{pmatrix} \times 10^{-10} \, .
\end{equation}
Then, a data vector with weak \ngy\ according to \eq{eq:fnldef} was
produced and superimposed with \gn\ white noise. Constructed in this
way, it is of the order $\mathcal{O}(10^{-5})$, thus the amplitude of
the resulting signal $s$ is comparable to CMB anisotropies.

The data vector had a length of $10^6$ pixels; for simplicity, we set
the beam function $B$ and the linear transformation matrix $M$ to
unity. This setup allows a brute force implementation of all equations
at a sufficient computational speed. We define the signal-to-noise
ratio ($S/N$) per pixel as the standard deviation of the input signal
divided by the standard deviation of the additive noise. It was chosen
in the range 0.5-10 to model the typical S/N per pixel of most CMB
experiments. To reconstruct the signal, we draw $1000$ samples
according to the scheme in \eq{eq:scheme}.

Whereas the $\Phi_{\NL}$ can be generated directly from a simple \gn\
distribution with known mean and variance, the construction of the
\fnl\ is slightly more complex. For each $\Phi_{\NL}$, we ran a
Metropolis Hastings algorithm with symmetric \gn\ proposal density
with a width comparable to that of the target density and started the
chain at $f_{\NL} = 0$. We run the \fnl\ chain to convergence. We
ensured that after ten accepted steps the sampler has decorrelated
from the starting point. Our tests conducted with several chains run
in parallel give $1 < R < 1.01$, where R is the convergence statistic
proposed by \citet{199211}. We record the last element of the chain as
the new \fnl\ sample.

Finally, we compared the obtained sets of values $\{\Phi_{\NL}^i\}$,
$\{f_{\NL}^i\}$ to the initial data. An example is shown in
\fig{fig:ex_phi_l}, where we illustrate the reconstruction of a given
potential $\Phi_{\NL}$ for different signal-to-noise ratios per
pixel. The $1-\sigma$ error bounds are calculated from the 16~\% and
84~\% quantile of the generated sample. Typical posterior densities
for \fnl\ as derived from the sample can be seen in
\fig{fig:ex_f_nl_1}. We considered the cases $f_{\NL} = 0$ and
$f_{\NL} = 200$ with $S/N = 10$ per pixel and show the distributions
generated from $1000$ draws. The derived posterior densities possesses
a mean value of $f_{\NL} = 6 \pm 40$ and $f_{\NL} = 201 \pm 40$,
respectively. The width of the posterior is determined by both the
shape of the conditional PDF of \fnl\ for a given $\Phi_{\NL}$ and the
shift of this distribution for different draws of $\Phi_{\NL}$
(\fig{fig:ex_f_nl_2}). The analysis of several data sets indicate that
the approximation does not bias the posterior density if the data are
decisive. We illustrate this issue in the left panel of
\fig{fig:f_nl_mean_error}, where we show the distribution of the mean
values $\langle f_{\NL} \rangle$ of the posterior density constructed
from 100 independent simulations. For an input value of $f_{\NL} =
200$ we derive a mean value $\langle f_{\NL} \rangle = 199.3 \pm 34.8$
and conclude that our sampler is unbiased for these input
parameters. For a high noise level, however, the $\Phi_{\NL}$ can
always be sampled such that they are purely \gn\ fields and thus the
resulting PDF for \fnl\ is then shifted towards $f_{\NL} = 0$. This
behavior is demonstrated in \fig{fig:f_nl_low_sn} where we compare the
constructed posterior density for the cases $S/N = 10$ and $S/N = 0.5$
per pixel. If the noise level becomes high, the approximated prior
distribution dominates and leads to both, a systematic displacement
and an artificially reduced width of the posterior. Therefore, the
sampler constructed here is conservative in a sense that it will tend
to underpredict the value of \fnl\ if the data are ambiguous.

An example of the evolution of the drawn \fnl\ samples with time can
be seen in \fig{fig:ex_f_nl_3}, where we in addition show the
corresponding autocorrelation function as defined via
\begin{equation}
  \xi (\Delta N) = \frac{1}{N} \sum^{N}_{i} \frac{(f_{\NL}^i - \mu)
    \cdot (f_{\NL}^{i + \Delta N} - \mu)}{\sigma^2} \, ,
\end{equation}
where N is the length of the generated \fnl\ chain with mean $\mu$ and
variance $\sigma^2$. The uncorrelated samples of \fnl\ ensure an
excellent mixing of the chain resulting in a fast convergence rate.

\section{Optimality}

In a frequentist analysis, parameter inference corresponds to finding
an estimator that enables to compute the most probable value of the
quantity of interest as well as a bound for the error. Ideally, the
estimator is unbiased and optimal, i.e.\ it's expectation value
coincides with the true value of the parameter and the error satisfies
the Cramer-Rao bound. Contrary, in a Bayesian approach, one
calculates the full probability distribution of the parameter
directly. Strictly speaking, optimality is therefore an ill-defined
term within the Bayesian framework. All we have to show is that the
approximation adopted in \eq{eq:jointdistr_nl} does not affect the
outcome of the calculation significantly. Note that the simplification
corresponds to imposing the prior of a purely \gn\ data set. In the
case of the CMB, this is a very reasonable assumption because up to
now no detection of \fnl\ has been reported.

To investigate the effects of the approximation, we checked the
dependence of the width of the posterior distribution on \fnl\ by
running a set of simulations with varying input values $f_{\NL} = 0$,
50, 100, 150, 200, 250. The estimated standard deviation
$\sigma_{f_{\NL}}$ of the drawn \fnl\ samples, each averaged over 10
simulation runs, are depicted in the right panel of
\fig{fig:f_nl_mean_error}. Contrary to the KSW estimator that shows an
increase of $\sigma_{f_{\NL}}$ with \fnl, we find no such indication
of sub-optimal behavior in the relevant region of small \ngy. In
particular, as the width of the distribution stays constant in the
limit $f_{\NL} \rightarrow 0$ where our approximated equations evolve
into the exact expressions, we conclude that the adopted
simplification does not affect the result significantly.

This finding can also be interpreted from a different point of view:
It is possible to define a frequentist estimator for \fnl\ based on
the mean of the posterior distribution. Our results indicate that such
an estimator is unbiased in the high signal-to-noise regime.

We apply an additional test in the next section where we compare our
sampling algorithm to a slower but exact scheme.

\twofig{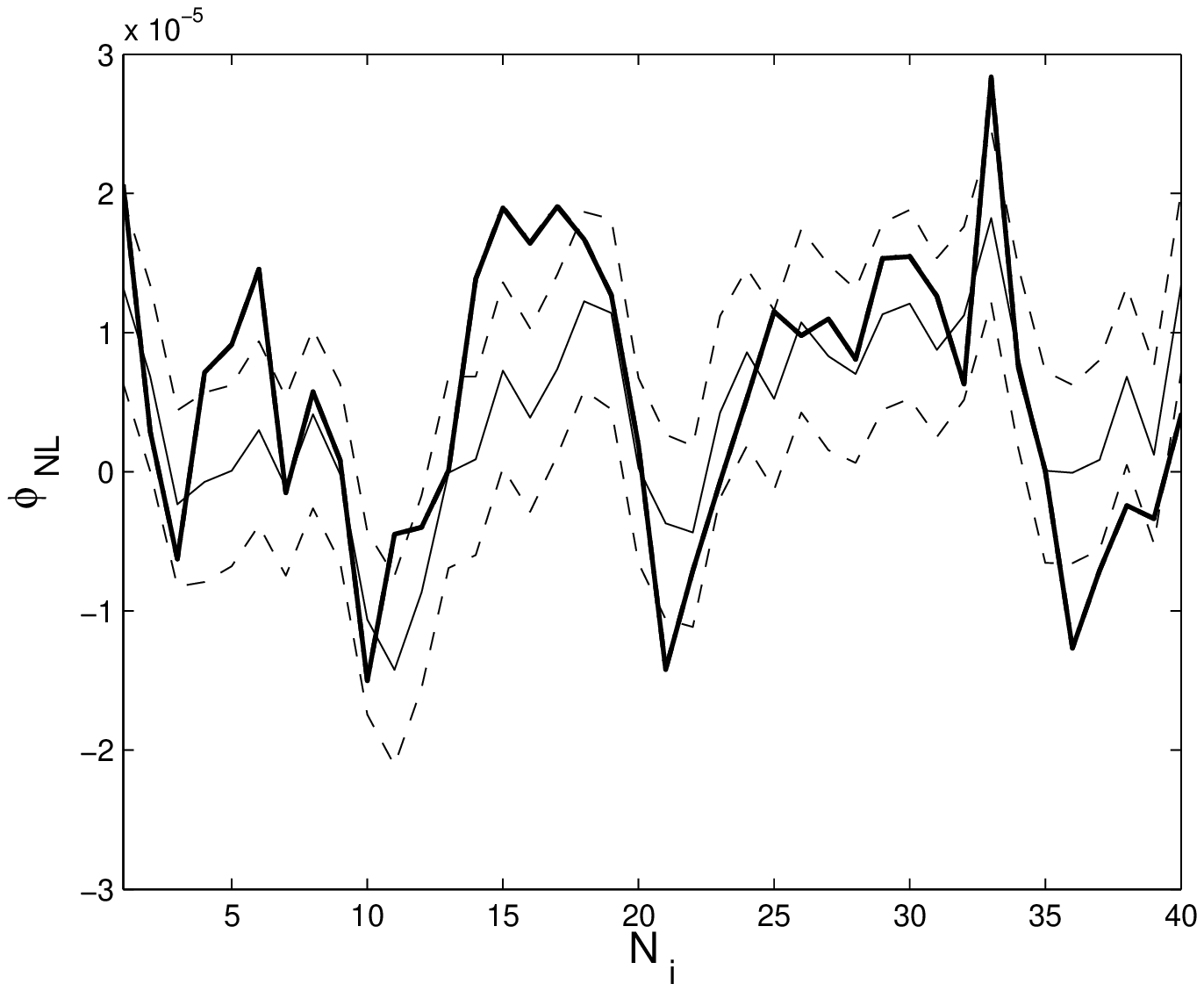}{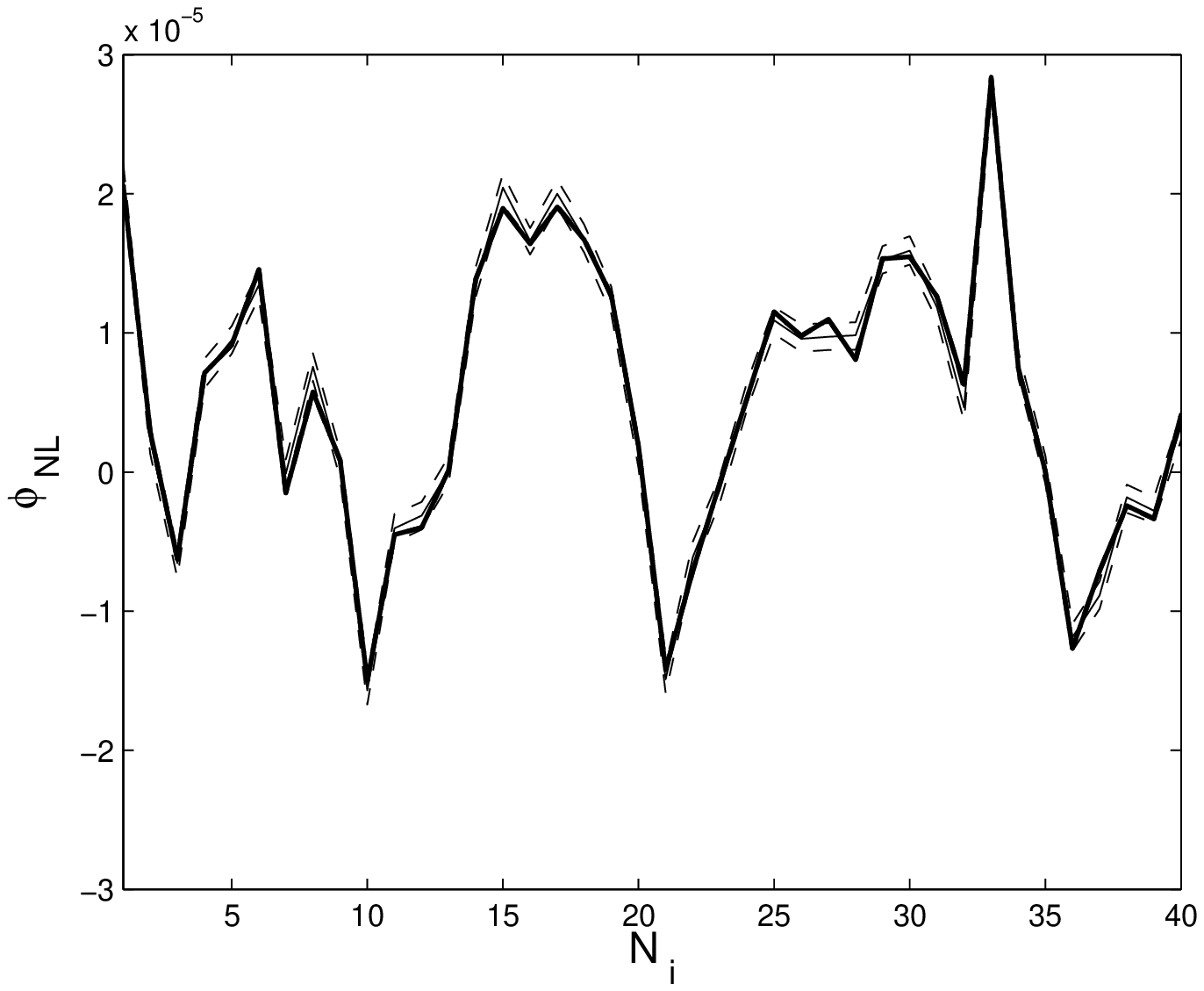}{fig:ex_phi_l}%
{Examples of reconstructed potentials $\Phi_{\NL}$. \emph{Left panel:}
  The input parameters for the calculation were $f_{\NL} = 200$ and
  $S/N = 1$. \emph{Right panel:} Analysis of the same data set for a
  signal-to-noise ratio of $S/N = 10$. For clarity, we show only 40
  elements of the $\Phi_{\NL}$-vector (\emph{thick solid line}) and
  its reconstruction (\emph{thin solid line}) as well as the
  $1-\sigma$ error bounds (\emph{dashed lines}). As the difference
  between $\Phi_{\NL}$ and the linear potential $\Phi_{\LL}$ is very
  small, $\Phi_{\LL}$ can not be distinguished from $\Phi_{\NL}$ in
  this plot. In both cases 1000 samples were drawn.}

\twofig{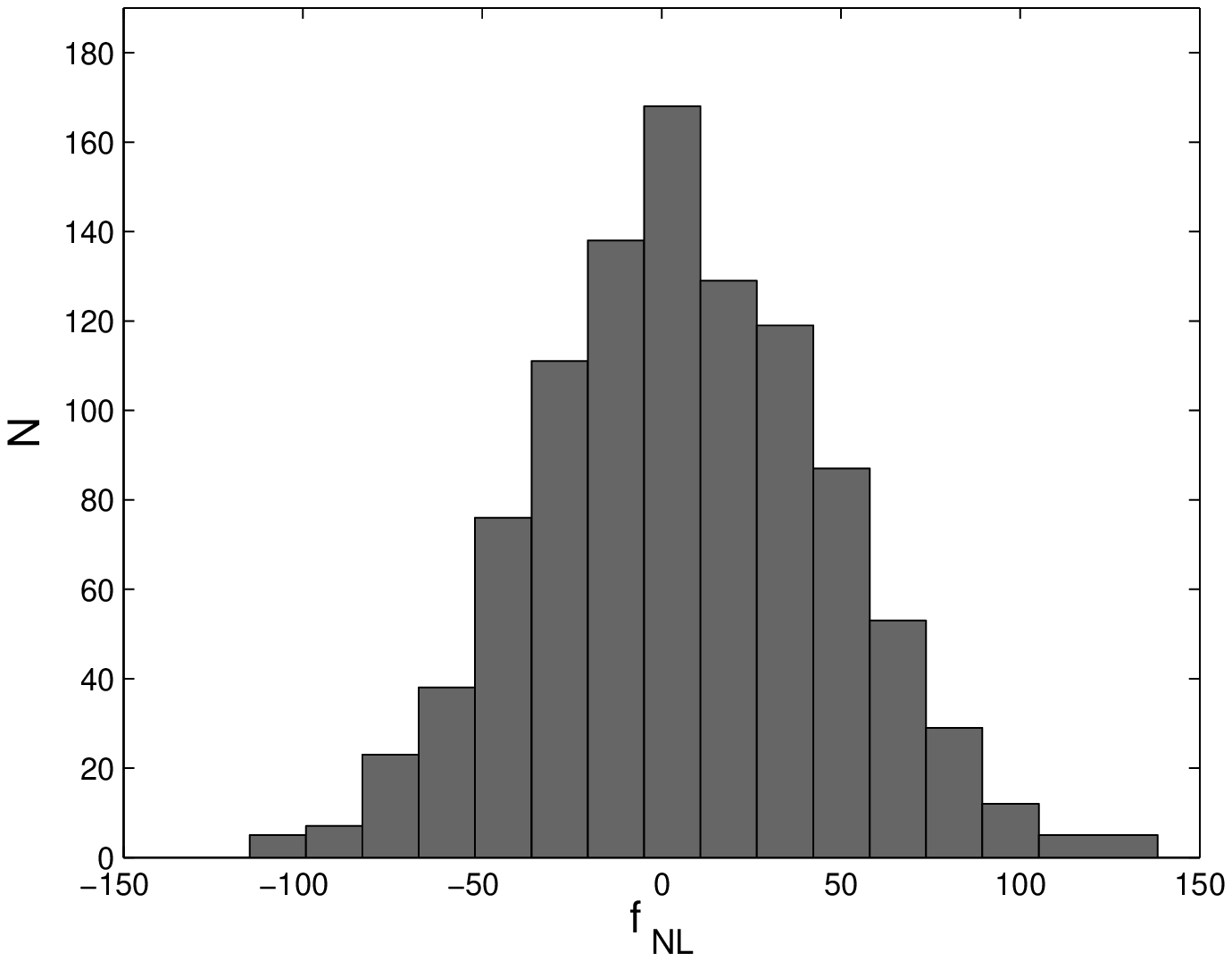}{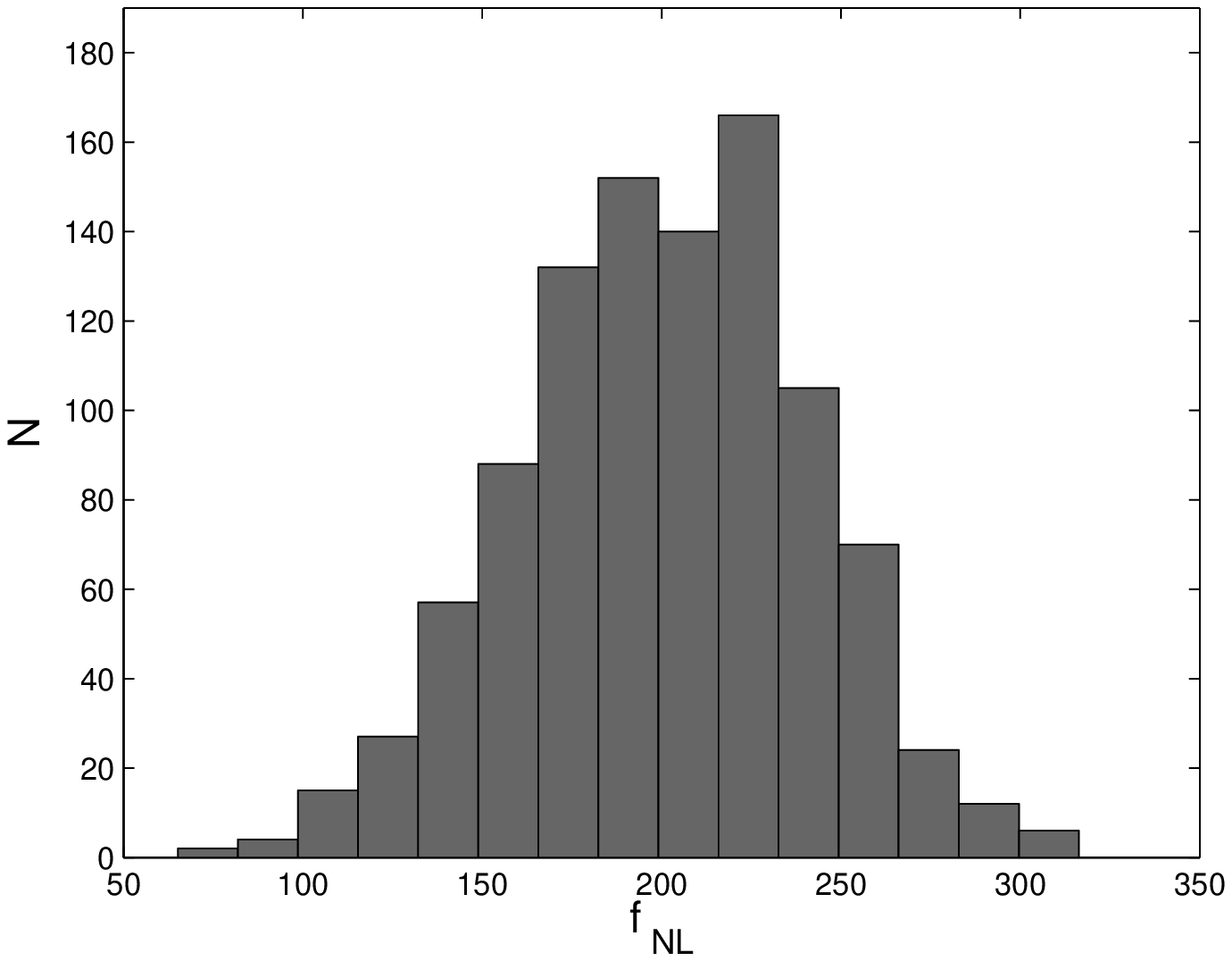}{fig:ex_f_nl_1}%
{Examples of a constructed posterior distribution for \fnl. The input
  parameters used in this runs were $N_{pix} = 10^6$, $S/N = 10$ and
  $f_{\NL} = 0$ (\emph{left panel}) or $f_{\NL} = 200$ (\emph{right
    panel}). For each parameter combination 1000 samples were drawn.}

\twofig{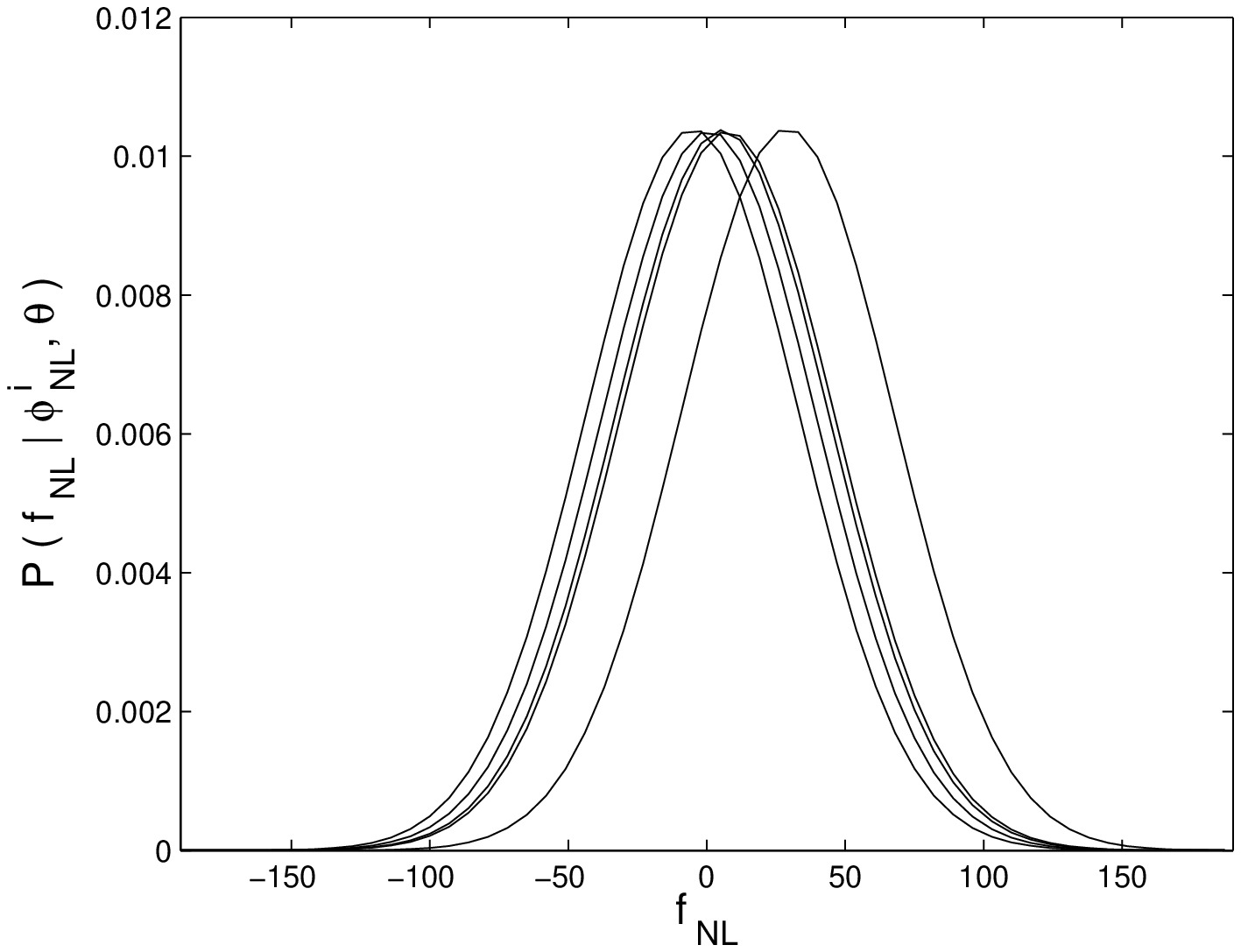}{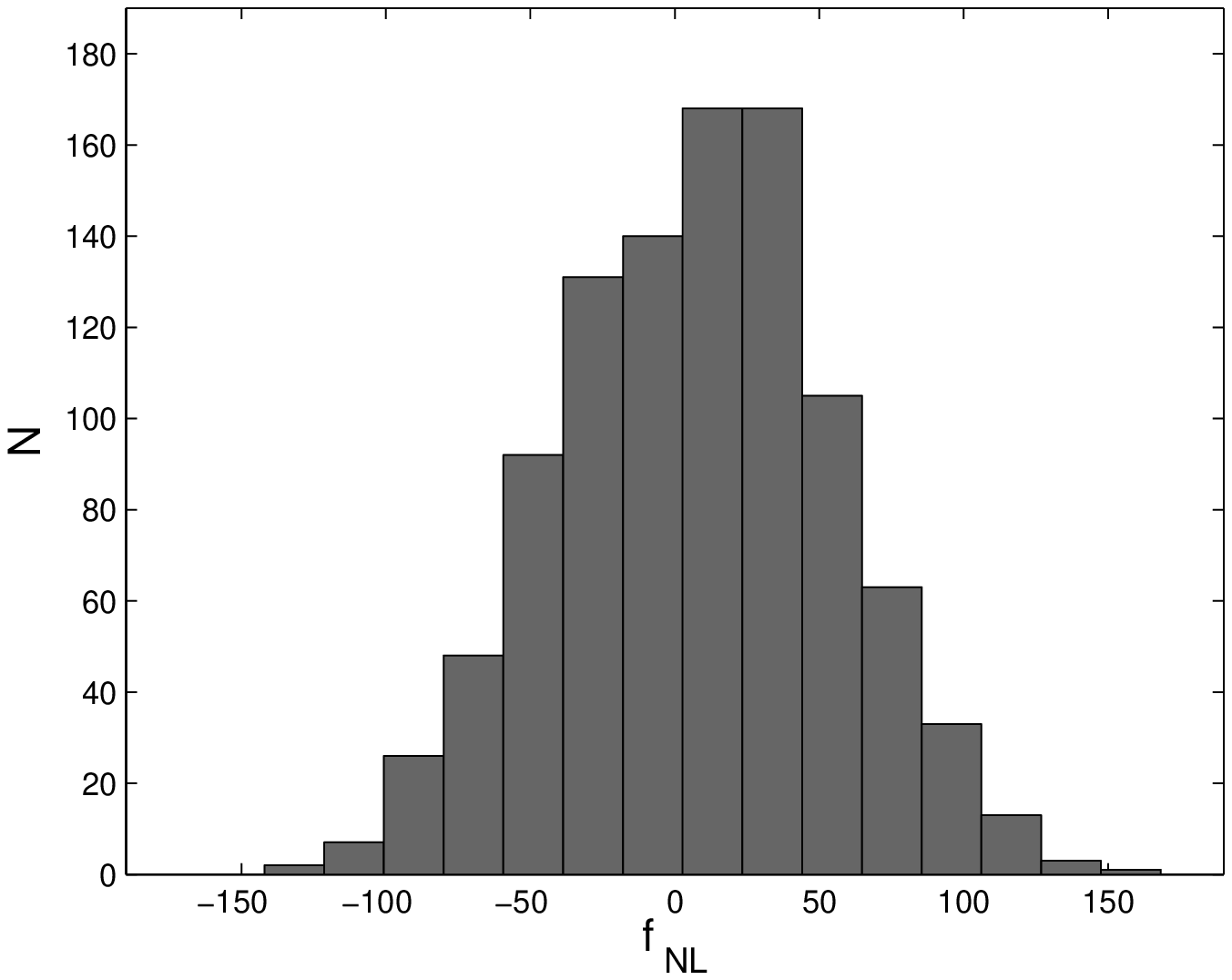}{fig:ex_f_nl_2}%
{Build-up of the posterior distribution of \fnl. We depict the
  conditional probability distributions $P(f_{\NL}| \Phi_{\NL},
  \theta)$ for several realizations of $\Phi_{\NL}$ (\emph{left
    panel}) and the constructed posterior after 1000 drawn samples
  (\emph{right panel}). The input parameters were chosen to be
  $N_{pix} = 10^6$, $f_{\NL} = 0$, and $S/N = 2$.}

\twofig{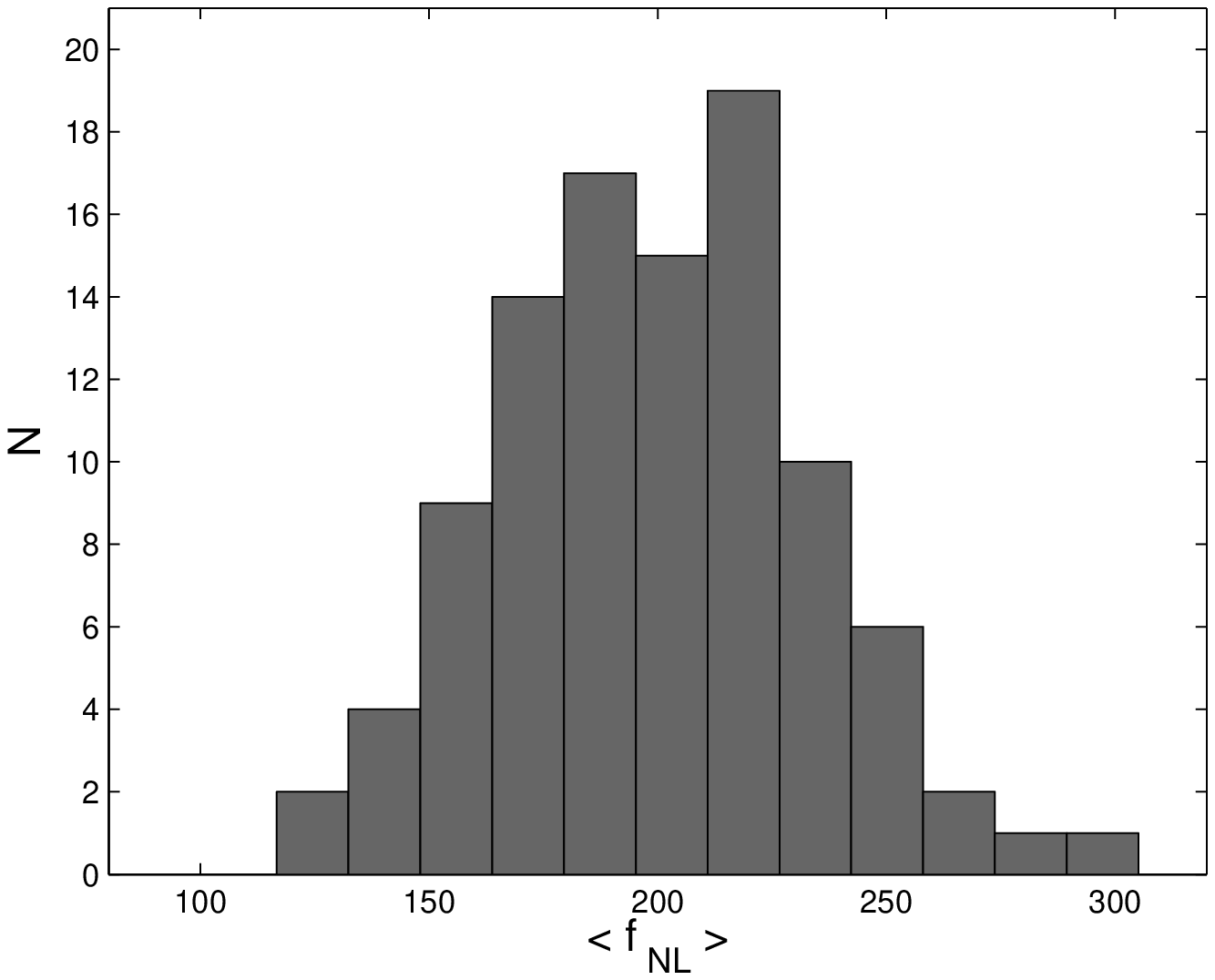}{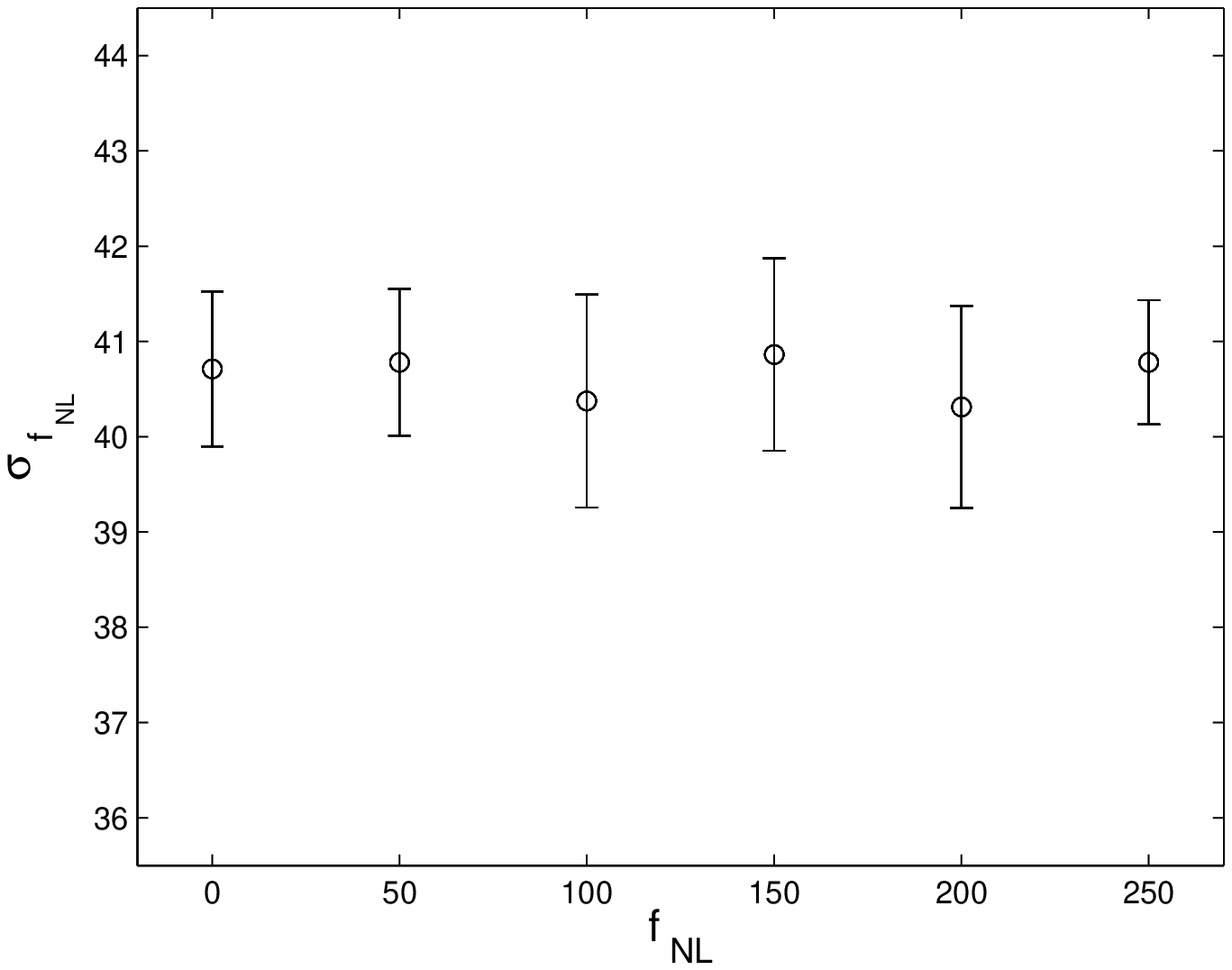}{fig:f_nl_mean_error}%
{Properties of the sampler. \emph{Left panel:} Shown is the
  distribution of the derived mean values of \fnl\ from 100
  simulations for a fiducial value of $f_{\NL} = 200$. \emph{Right
    panel:} We display the estimated standard deviation
  $\sigma_{f_{\NL}}$ of the drawn \fnl\ samples as a function of
  \fnl. Each data point is averaged over 10 simulations. The input
  parameters used in this runs were $N_{pix} = 10^6$ and $S/N = 10$,
  in each simulation 1000 samples were drawn.}

\threefig{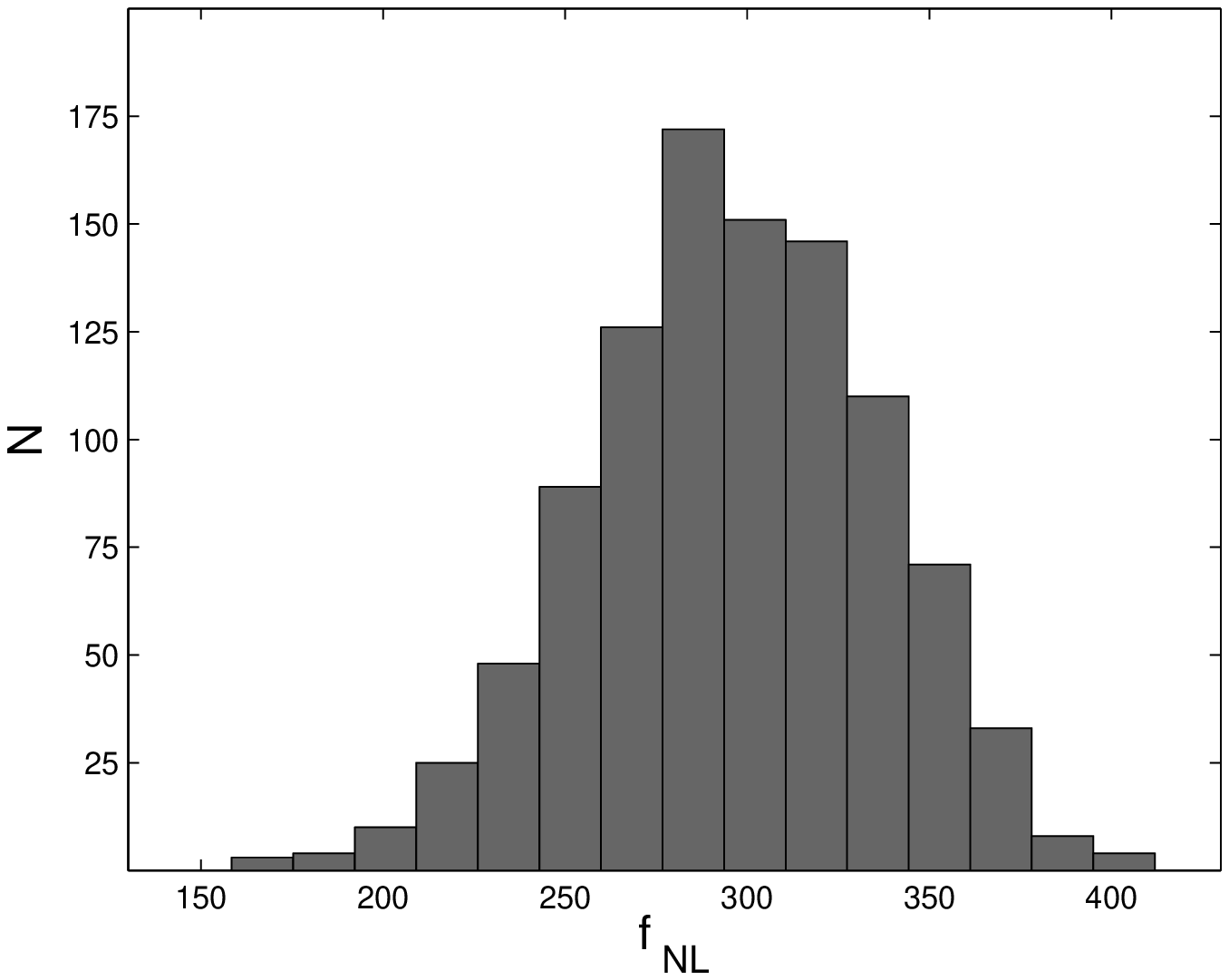}{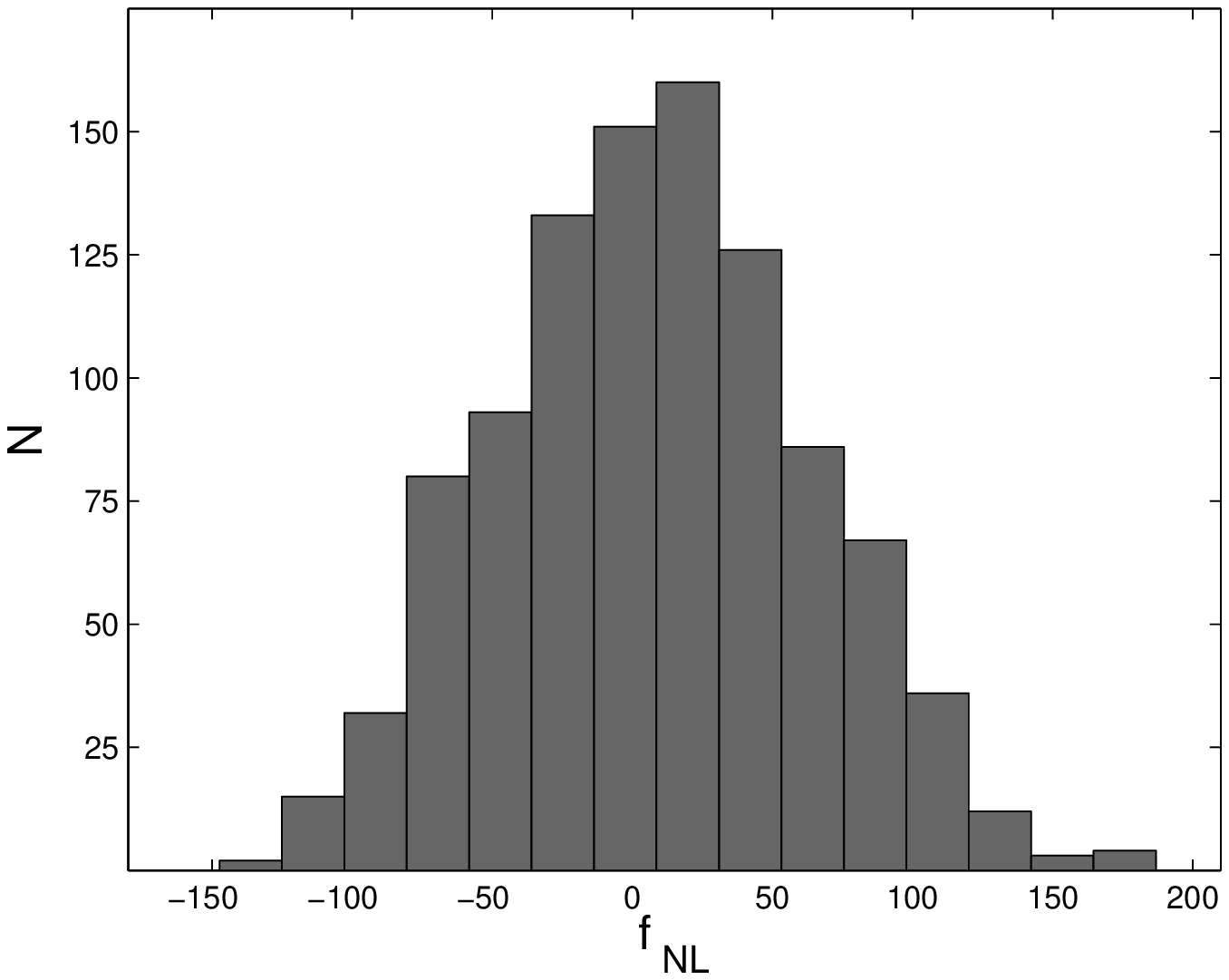}{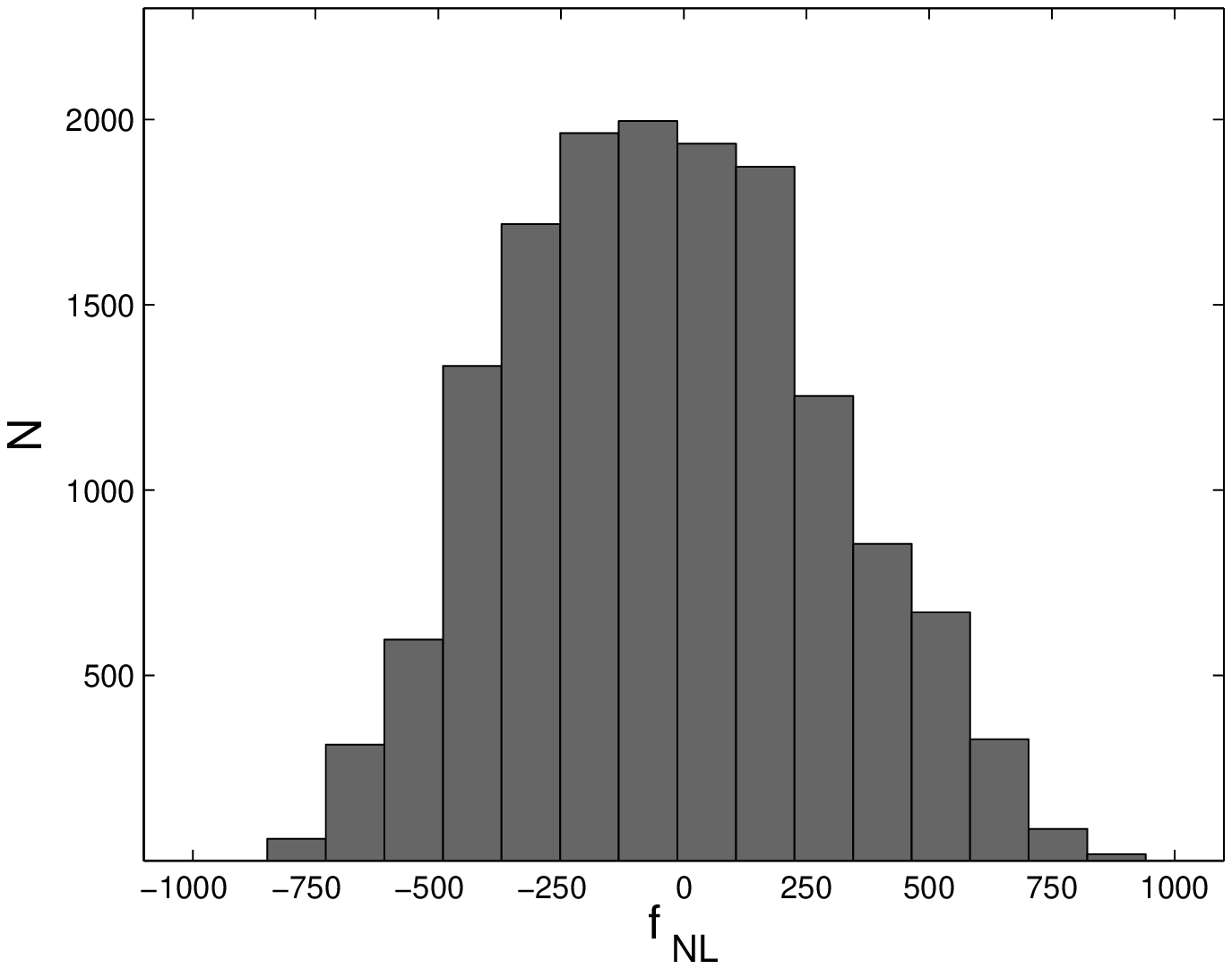}{fig:f_nl_low_sn}%
{Impact of the signal-to-noise ratio on the approximate sampling
  scheme. \emph{Left panel:} Example of a constructed posterior
  distribution for $S/N = 10$. \emph{Middle panel:} Analysis of the
  same data set, but for $S/N = 0.5$. At high noise level, the
  distribution becomes too narrow and systematically shifted towards
  $f_{\NL} = 0$. \emph{Right panel:} For comparison, we show the
  analysis of the data set at $S/N = 0.5$ using exact Hamiltonian
  Monte Carlo sampling. As input parameters, we used $f_{\NL} = 300$
  and $N_{pix} = 10^6$. For the approximate and exact analysis, 1000
  and $15\,000$ samples were drawn, respectively.}

\twofig{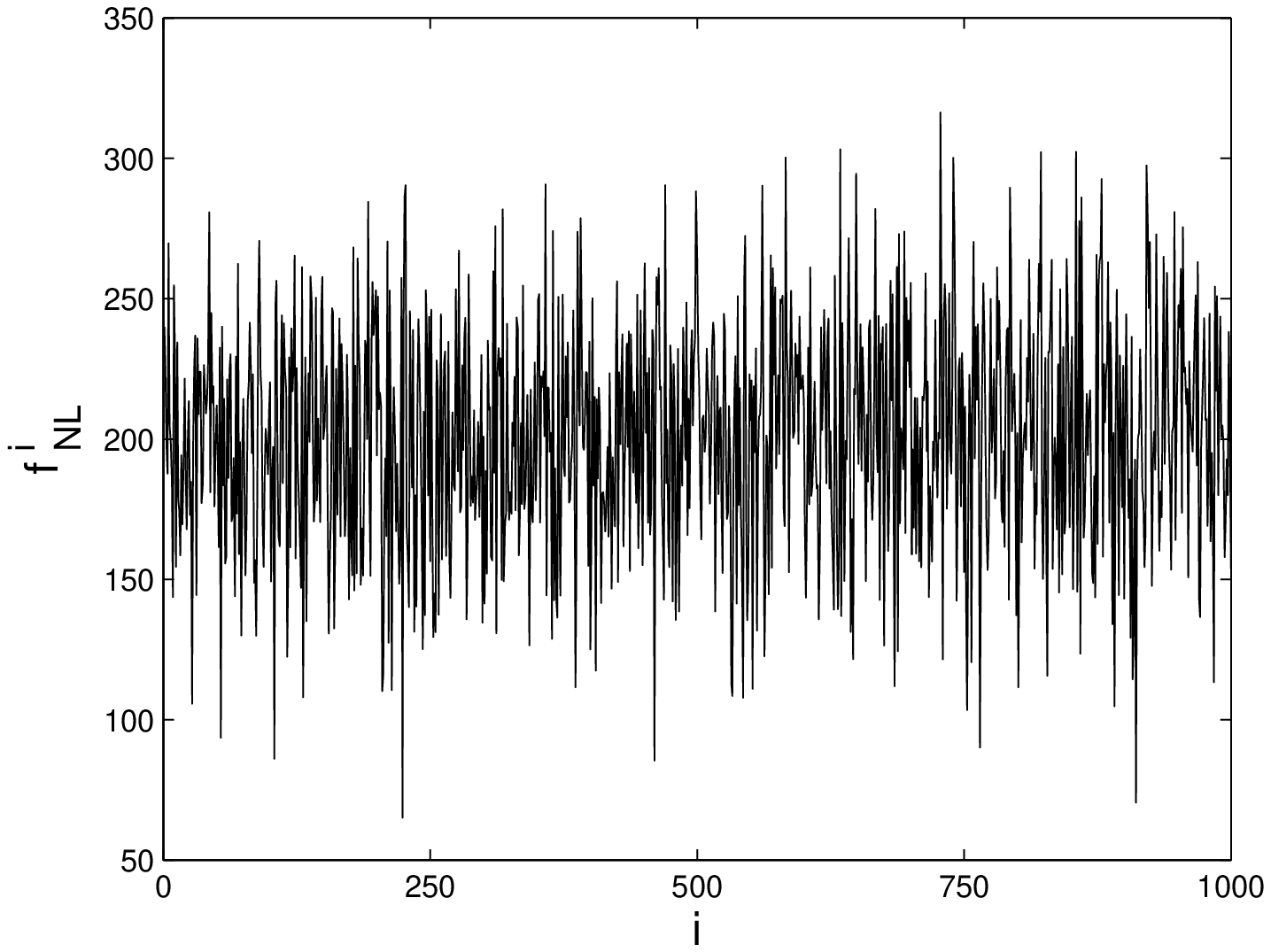}{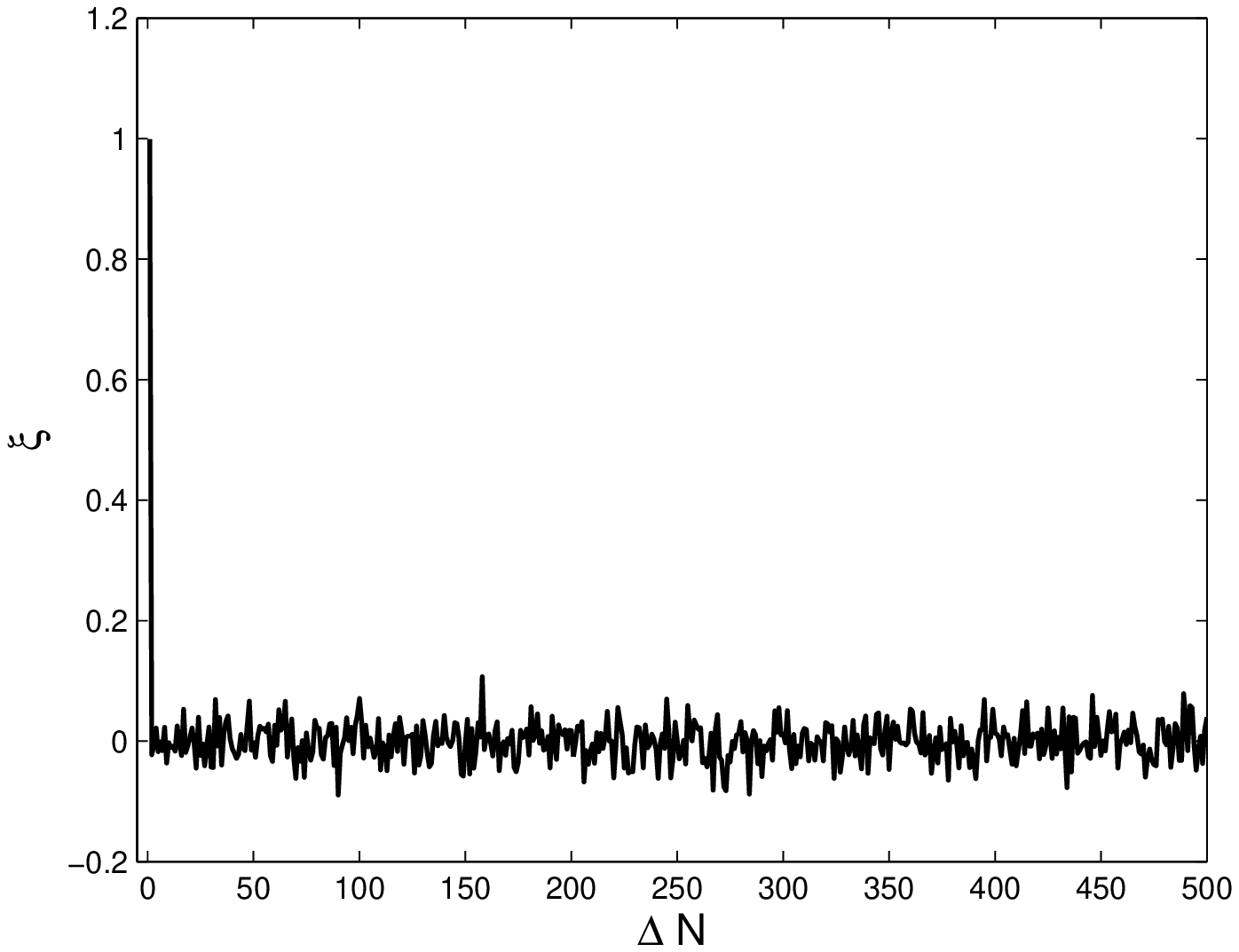}{fig:ex_f_nl_3}%
{Example ${f_{\NL}}$ chain. \emph{Left panel:} We display the chain of
  1000 generated \fnl\ samples which built up the histogram plotted on
  the right hand side in \fig{fig:ex_f_nl_1}. \emph{Right panel:} The
  autocorrelation function of \fnl\ confirms the uncorrelatedness of
  the samples.}

\section{Hamiltonian Monte Carlo sampling}
\label{sec:hmc}

In addition to the sampling technique presented above, we tested
whether an exact Hamiltonian Monte Carlo (HMC) sampler is applicable
to the problem. Within this approach one uses the methods developed in
classical mechanics to describe the motion of particles in
potentials. The quantity of interest is regarded as the spatial
coordinate of a particle and the potential well corresponds to the PDF
to evaluate \citep{Duane1987}. To each variable $(f_{\NL},
\Phi_{{\LL}, 1},\dots,\Phi_{{\LL}, n})$, a mass and a momentum is
assigned and the system is evolved deterministically from a starting
point according to the Hamilton equations of motion.

The applicability of HMC sampling techniques to cosmological parameter
estimation has been demonstrated in \citet{2007PhRvD..75h3525H}, and
the authors of \citet{2008MNRAS.389.1284T} compared HMC with Gibbs
sampling for CMB power spectrum analysis. To apply HMC sampling to
\fnl\ inference, we deduced the expression of the Hamiltonian

\begin{equation}
  H = \sum_{i} \frac{p^2_{i}}{2 \, m_{i}} - \log[
  P(d,\Phi_{\LL},f_{\NL},\theta) ] \, ,
\end{equation}
where the potential is related to the PDF as defined in
\eq{eq:jointdistr_l}. The Hamilton equations of motion,
\begin{align}
\frac{{\mathrm{d}}x_{i}}{\mathrm{dt}} &= \frac{\partial H}{\partial
  p_{i}} \, , \nonumber \\ \frac{{\mathrm{d}}p_{i}}{\mathrm{dt}} &=
-\frac{\partial H}{\partial {x_{i}}} = \frac{\partial
  \log[P(d,\Phi_{\LL},f_{\NL},\theta)]}{\partial {x_{i}}} \, ,
\end{align}
are integrated for each parameter $\{x_{i}; \ p_{i}\} = \{f_{\NL}, \
\Phi_{\LL}; \ p_{f_{\NL}}, \ p_{\Phi_{\LL}}\}$ using the leapfrog
method with step size $\delta t$,
\begin{align}
  p_{i}(t + \frac{\delta t}{2}) &= p_{i}(t) + \frac{\delta t}{2} \,
  \frac{\partial \log[P(d,\Phi_{\LL},f_{\NL},\theta)]}{\partial x_{i}}
  \bigg|_{x(t)} \nonumber \\ x_{i}(t + \delta t) &= x_{i}(t) +
  \frac{\delta t}{m_i} \, p_{i}(t + \frac{\delta t}{2}) \nonumber \\
  p_{i}(t + \delta t) &= p_{i}(t + \frac{\delta t}{2}) + \frac{\delta
    t}{2} \, \frac{\partial
    \log[P(d,\Phi_{\LL},f_{\NL},\theta)]}{\partial {x_{i}}}\bigg|_{x(t
    + \delta t)} \, .
\end{align}
The equations of motion for $x_{i}$ are straightforward to compute, as
they only depend on the momentum variable. To integrate the evolution
equations for the $p_{i}$, we derive
\begin{align}
\label{eq:hmc_grad}
\frac{\partial \log[P(d,\Phi_{\LL},f_{\NL},\theta)]}{\partial f_{\NL}}
&= (\Phi^2_{\LL} - \langle \Phi^2_{\LL} \rangle)^{\dagger} M^{\dagger}
B^{\dagger} N^{-1} (d - B M \Phi_{\LL} - f_{\NL} B M (\Phi^2_{\LL} -
\langle \Phi^2_{\LL} \rangle)) \, , \nonumber \\ \frac{\partial
  \log[P(d,\Phi_{\LL},f_{\NL},\theta)]}{\partial {\Phi_{\LL}}}
&\approx M^{\dagger} B^{\dagger} N^{-1}(d - B M \Phi_{\LL}) -
P_\Phi^{-1} \Phi_{\LL} + \ 2 f_{\NL} \operatorname{diag}(M^{\dagger}
B^{\dagger} N^{-1} d) \Phi_{\LL} + \mathcal{O}(\Phi_{\LL}^2) \, ,
\end{align}
where we have truncated the gradient in the latter equation at order
$\mathcal{O}(\Phi_{\LL}^2)$. The final point of the trajectory is
accepted with probability $p = \min(1, \exp[-\Delta H ])$, where
$\Delta H$ is the difference in energy between the end- and starting
point. This accept/reject step allows us to restore exactness as it
eliminates the error introduced by approximating the gradient in
\eq{eq:hmc_grad} and by the numerical integration scheme. In general,
only accurate integrations where $\Delta H$ is close to zero result in
high acceptance rates. Furthermore, the efficiency of a HMC sampler is
sensitive to the choice of the free parameters $m_i$, which
corresponds to a mass. This issue is of particular importance if the
quantities of interest possess variances varying by orders of
magnitude. Following \citet{2008MNRAS.389.1284T}, we chose the masses
inversely proportional to the diagonal elements of the covariance
matrix which we reconstructed out of the solution of the sampling
scheme from \sect{sec:sampling}. We initialized the algorithm by
performing one draw of $\Phi_{\NL}$ from the conditional PDF
$P(\Phi_{\NL}|d,\theta)$ and setting $f_{\NL} = 0$. The outcome of
repeated analyses of the data set presented in \fig{fig:ex_f_nl_2} is
shown in \fig{fig:hmc}. The consistency of the distributions confirms
the equivalence of the two sampling techniques in the high
signal-to-noise regime. However, convergence for the HMC is far
slower, even for the idealized choice for $m_i$ and a reasonable
starting guess, as can be seen from the large width of the
autocorrelation function (see right panel of \fig{fig:hmc}).

We conclude, therefore, that the direct sampling scheme presented in
\sect{sec:sampling} is more efficient than HMC when applied to the
detection of local \ngs\ in the high signal-to-noise
regime. However, as shown in the rightmost panel of
\fig{fig:f_nl_low_sn}, the exact analysis using a HMC algorithm
remains applicable at high noise level.

\twofig{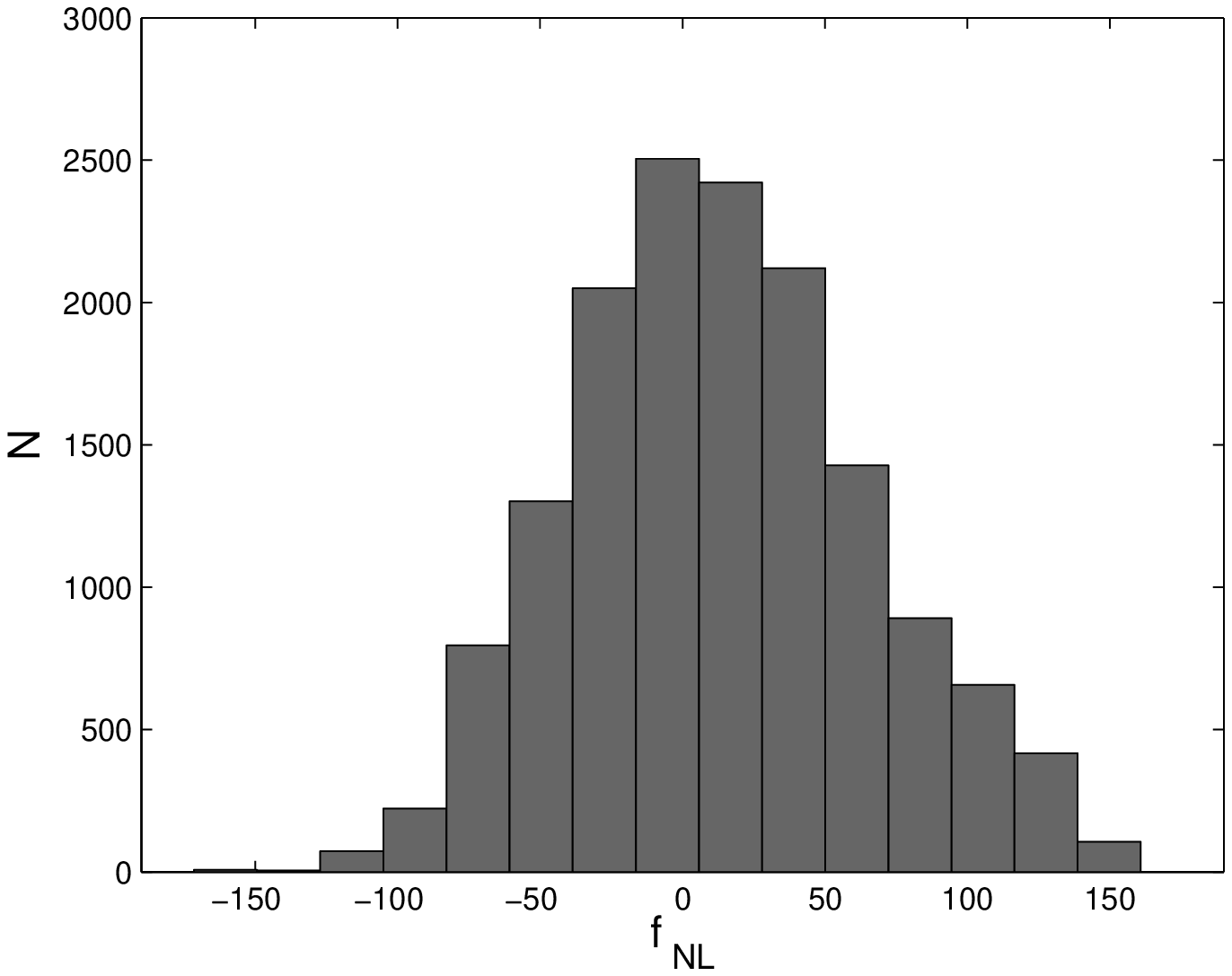}{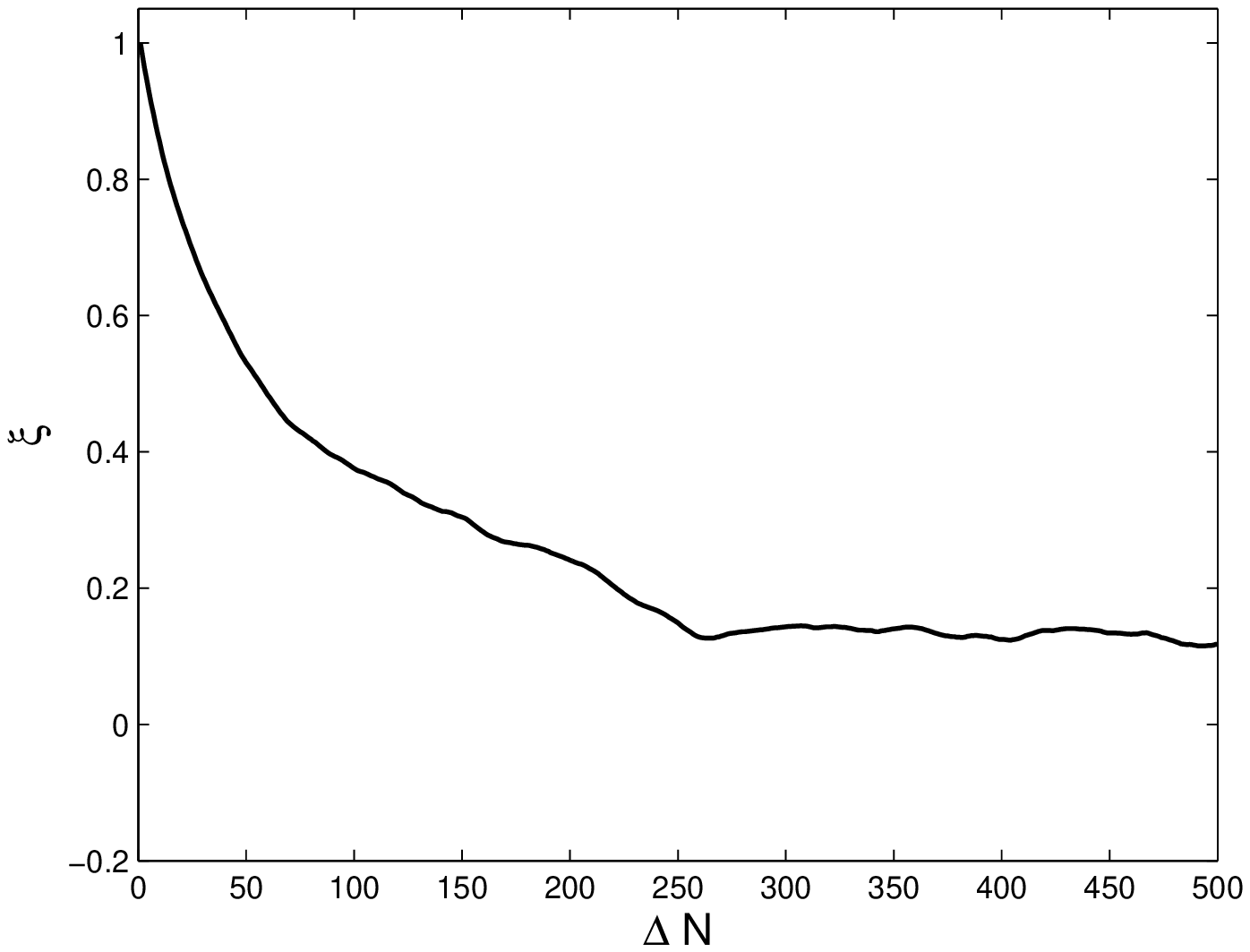}{fig:hmc}%
{Performance of the Hamilton Monte Carlo sampler. \emph{Left panel}:
  Analysis of the data set of \fig{fig:ex_f_nl_2} using the HMC
  sampler. Here, $15\,000$ samples were draw. \emph{Right panel}: The
  autocorrelation function of \fnl.}

\section{Extension to realistic data}
\label{sec:cmb}

Applying the method to a realistic CMB data set requires recovering
the primordial potential $\Phi_{\LL}$ on shells at numerous distances
$r_i$ from the origin to the present time cosmic horizon. Thus, the
product of the transfer matrix $M$ with the potential transforms to
\begin{equation}
M \Phi \rightarrow \sum_{i} M_{i} \Phi(r_i) \, .
\end{equation}
Here, the matrix $M$ projects a weighted combination of the
$\Phi(r_i)$ at different radii to a resulting two dimensional signal
map $s$. The resolution of the $r$-grid can be coarser where the
transfer functions for radiation are close to zero and must be finer
at the distances of recombination and reionization. Another
modification concerns the covariance matrix $P_{\Phi}$ of the
potential. Now it additionally describes the correlation of $\Phi$ on
distinct shells at different distances,
\begin{equation}
  \Phi^{\dagger} P_\Phi^{-1} \Phi \rightarrow \sum_{i, j} \Phi(r_i)
  P^{-1}_{\Phi(r_i), \Phi(r_j)} \Phi(r_j) \, ,
\end{equation}
and can be calculated from the primordial power spectrum
$\mathcal{P}(k)$ predicted by inflation
\begin{equation}
  P_{\Phi(r_i), \Phi(r_j) \; \ell} = \frac{2}{\pi} \, \int k^2 dk \,
  \mathcal{P}(k) \, j_\ell(k\,r_1) \, j_\ell(k\,r_2) \, .
\end{equation}

To tighten the constraints on $\Phi$, polarization information can be
included into the analysis as well simply by replacing the temperature
by the polarization transfer function in the expression for $M$. We
plan to study the application of our methods to realistic CMB data in
a future publication.

The computational speed of a complete implementation is limited by
harmonic transforms that scale as $\mathcal{O}(N_{pix}^{3/2})$.

\section{Summary}
\label{sec:summary}

In this paper, we developed two methods to infer the amplitude of the
\ngy\ parameter \fnl\ from a data set within a Bayesian approach. We
focused on the so called local type of \ngy\ and derived an expression
for the joint probability distribution of \fnl\ and the primordial
curvature perturbations, $\Phi$. Despite the methods are of general
validity, we tailored our discussion to the case example of CMB data
analysis.

We developed an exact Markov Chain sampler that generates correlated
samples from the joint density using the Hamiltonian Monte Carlo
approach. We implemented the HMC sampler and applied it to a toy model
consisting of simulated measurements of a 1-D sky. These simulations
demonstrate that the recovered posterior distribution is consistent
with the level of simulated \ngy.

With two approximations that exploit the fact that the \ngn\
contribution to the signal is next order in perturbation theory, we
find a far more computationally efficient Monte Carlo sampling
algorithm that produces \emph{independent} samples from the \fnl\
posterior. The regime of applicability for this approximation is for
data with high signal-to-noise and weak \ngy.

By comparison to the exact HMC sampler, we show that our approximate
algorithm reproduces the posterior location and shape in its regime of
applicability. If non-zero \fnl\ is not supported by the data the
method is biased towards Gaussianity. The approximate posterior more
strongly prefers zero \fnl\ compared to non-zero values than the exact
posterior, as expected given the nature of the approximations which
Gaussianize the prior. This method is therefore only applicable if the
data contains sufficient support for the presence of \ngy\,
essentially overruling the preference for Gaussianity in our
approximate prior.

Our efficient method enables us to perform a Monte Carlo study of the
behavior of the posterior density for our toy model data with high
signal-to-noise per pixel. We found that the width of the posterior
distribution does not change as a function of the level of \ngy\ in
the data, contrary to the frequentist estimator where there is an
additional, \fnl\ dependent, variance component
\citep{2007JCAP...03..019C, 2007PhRvD..76j5016L}. Our results suggest
that this may be an advantage of the Bayesian approach compared to the
frequentist approach, motivating further study of the application of
Bayesian statistics to the search for primordial local \ngy\ in
current and future CMB data.

We close on a somewhat philosophical remark. Even though we chose a
Gaussian prior approximation for expediency, it may actually be an
accurate model of prior belief for many cosmologists since canonical
theoretical models predict Gaussian perturbations. From that
perspective our fast, approximate method may offer some
(philosophically interesting) insight into the question ``what level
of signal-to-noise in the data is required to convince someone of the
presence of \ngy\ whose prior belief is that the primordial
perturbations are Gaussian?''

\begin{acknowledgements}
  We thank the anonymous referee for the comments which helped to
  improve the presentation of our results. We are grateful to Rob
  Tandy for doing initial tests on detecting primordial \ngy\ in
  reconstructed sky maps. We thank Anthony J. Banday for useful
  conversations. BDW is partially supported by NSF grants AST 0507676
  and AST 07-08849. BDW gratefully acknowledges the Alexander
  v. Humboldt Foundation's Friedrich Wilhelm Bessel Award which funded
  part of this work.
\end{acknowledgements}

\bibliographystyle{aa}
\bibliography{literature}

\end{document}